# A COMPARISON BETWEEN 1.5μm PHOTOLUMINESCENCE FROM Er-DOPED Si-RICH SiO$_2$ FILMS AND (Er,Ge) CO-DOPED SiO$_2$ FILMS

*J. Mayandi[1], T. G. Finstad[1, 2, a], C. L. Heng[1], Y. J. Li[2], A. Thøgersen[2], S. Foss[1] and H. Klette[3]*

[1] Dept. Physics, University of Oslo, Norway
[2] Ctr. Mater. Sci. & Nanotechnology, University of Oslo, Norway
[3] SINTEF, Mater. & Chem. Oslo, Norway

**ABSTRACT**

We have studied the 1.5 μm photoluminescence (PL) from Er ions after annealing two different sample sets in the temperature range 500 °C to 1100 °C. The different sample sets were made by magnetron sputtering from composite targets of Si+SiO$_2$+Er and Ge+SiO$_2$+Er respectively for the different sample sets. The annealing induces Si – and Ge-nanoclusters respectively in the different film sets. The PL peak reaches its maximum intensity after annealing at 700 °C for samples with Ge nanoclusters and after annealing at 800 °C for samples with Si. No luminescence from nanoclusters was detected in neither sample sets. This is interpreted as an energy transfer from the nanocluster to Er atoms. Transmission electron microscopy shows that after annealing to the respective temperature yielding the maximum PL intensity both the Ge and Si clusters are non-crystalline. Here we mainly compare the spectral shape of Er luminescence emitted in these different nanostructured matrixes. The PL spectral shapes are clearly different and witness a different local environment for the Er ions.

## 1. INTRODUCTION

Er$^{3+}$ has attracted great interest in the photonic technology, due to that the emitted wavelength of the $^4I_{13/2}$ to $^4I_{15/2}$ transition coincides with the minimum attenuation of silica-based optical fibers [1] and it finds use in optical amplifiers. There is a need for a cheap and more compact alternative to the present fiber optical amplifier especially to match an increasing need for signal transfer over relatively short distances. The approaches taken for future technological solutions to this overlap those undertaken to meet the need for a faster signal transfer between functional blocks on a computer chip, between chips and systems. Application motivated research on Er in Si has a long history that has foreseen the same need we see today. However luminescence from Er in Si has several problems in particular with two major de-excitation processes: (i) the phonon-assisted energy back-transfer originating from the near resonance between the Er-induced level in the band gap and the internal Er levels [2]; and (ii) the Auger-type interaction involving free carriers and excited Er ions [3]. The former induces a large temperature quenching in the photoluminescence (PL) intensity, while the latter increases with the density of excited carriers and gives rise also to an energy backflow mechanism. To remove or reduce these detrimental effects, a promising approach is currently explored through the widening of the band gap by using SiO$_2$ as a host material co-doped with Er ions and Si nanoclusters (Si-nc) which are very efficient sensitizers for Er luminescence with a two orders of magnitude increase in efficiency as first reported by Kenyon [4]. In order to improve on this type of structures it is necessary to understand the systems on the nanometer scale, both regarding the energy transfer processes occurring, their physical characteristics as well as how the systems can self-assemble on the nanometer scale and how this possibly can be controlled. A large number of scientists around the world have engaged in these problems of nanoscience. The simplified main elements of the resulting models for photoluminescence in such systems, consist of first absorption of electromagnetic radiation by the nanocrystal (or other nanostructures), which gets exited. Excitation energy is then transferred to the Er atom. The Er atom decays to the 1st exited level and then emits l.5 μm radiation. The absorption properties of the nanocrystals depends upon their size by quantum confinement of the electrons. The nanocrystal energy levels will also influence which levels the Er can be exited to in the transfer process. It is desired that the upper exited levels gets populated followed by a rapid decay down to the first exited level. The energy levels of the Si nanocrystal also determine the probability for a transfer of energy back from the Er ion to the Si nanocrystal, which is undesirable. The energy transfer process depends strongly on the interaction distance between the Er ion and Si nanocrystals. The interaction can be caused by direct wave function overlap or by dipole-dipole interactions. The Er atom needs to be in an environment making it optically active. An isolated Er atom is not optically active and likewise an Er atom bonded in an Er silicide is considered optically inactive. The Er atom is inactive when the luminescence efficiency is low, which can be caused by the radiative transition being much less probable than competing de-excitation mechanisms. The details in the models sketched are generally not known and they are hard to probe directly. Ge offers an interesting alternative to Si. Ge has many of the same chemical and material properties as Si and can be expected to behave in similar ways, while there are some important differences. A study of Ge nanostuctures in $SiO_2$ doped with Er can thus shine light on the situation with silicon and our group have undertaken such a study [5]. One interesting difference between Si and Ge in this respect is that Ge theoretically shows a stronger quantum confinement, due to the larger exciton Bohr radius, for equal size of quantum dots. The atomic mobilities in $SiO_2$ are larger than that of Si which makes it possible to make equal size nanoclusters at lower temperatures with Ge than with Si. This is an advantage for the tendency to Er clustering at increasing temperatures, which could leave Er optically inactive. Here we mainly intend to compare the spectral shape of the luminescence from Er with Ge and Si nanoclusters, respectively, in $SiO_2$ matrixes and use that as a probe for the local environments of Er.

## 2. EXPERIMENTAL DETAILS

The (Er, Ge) co-doped $SiO_2$ film was deposited on a low resistivity (~ $10^{-2}$ Ωcm) n-type Si (100) substrate. Before sputtering, the wafer was cleaned using a standard RCA procedure and given a dip in a 10% HF solution to remove the native oxide. Then the film was deposited on the substrates by sputtering an Er+Ge+$SiO_2$ composite target at 300 W. The base pressure of the chamber was less than $5 \times 10^{-7}$ Torr. The area ratios of the metal Er, Ge plates, and $SiO_2$ base in the target was 1: 3.3: 95.7. The working gas was Ar plasma with a gas pressure of 5.2 $\times 10^{-3}$ Torr. This structure was labeled sample type A and had a film thickness of 1.1 μm. For the preparation of the device with Si ncs we used a composite target with the ratio of 1:17:82 (Er: Si: $SiO_2$) and this structure is labeled as type B with a film thickness of about 25 nm on a p- type Si(100) substrate. (The structure was originally designed for electroluminescence MOS devices). Here we report photoluminescence (PL) measurements which were carried out by pumping with a 488-nm line of an Ar laser, which had a power of 100 mW and was focused to a beam spot of $6 \times 10^{-2}$ $cm^2$. The laser beam was chopped at a frequency of 72.5 Hz. A spectrometer consisting of a single-grating monochromator, a liquid-nitrogen-cooled Ge detector, and a lock-in amplifier was used in the PL measurements.

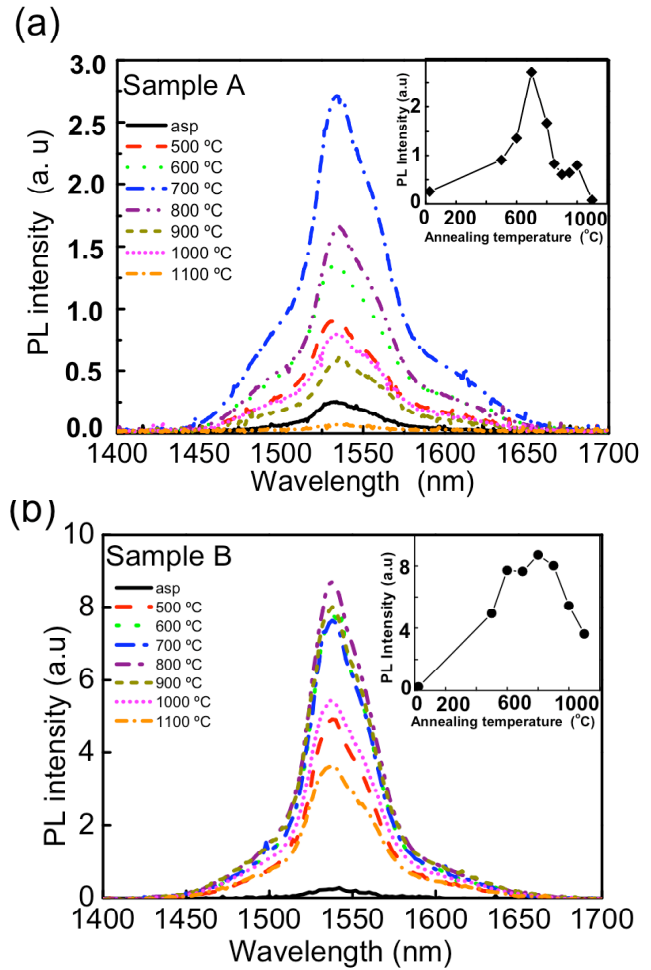

*FIG 1 The ~1.54 μm PL from the sample (a) type A (Er+Ge+$SiO_2$) and (b) type B (Er+Si+$SiO_2$). All the structures have been annealed at the range of 500 °C - 1100 °C in $N_2$ for 30 min. The insets in the graphs shows the PL intensity vs annealing temperature.*

## 3. RESULTS AND DISCUSSION

The annealing dependence of Er PL from both sample sets (A and B) has been studied for annealing temperatures from 500 °C to 1100 °C in $N_2$. The PL spectra are shown in Figs. 1(a) and 1(b), for sample set A and B respectively. The peak is attributed to the 4f inner-shell transition $^4I_{13/2} \rightarrow {}^4I_{15/2}$ transition of the $Er^{3+}$ as no signal above the noise level have been observed in control samples without Er prepared in previous experiments in this spectral range. The insets in Fig. 1 (a) and Fig. 1 (b) show how the PL peak intensity varies with the annealing temperature; it is relatively weak for the as-prepared samples before annealing for both sample sets; in both cases it increases with annealing temperature, goes through a maximum and then decrease for the highest temperatures. In Fig. 1(a), the $Er^{3+}$ PL peak intensity reaches a maximum after 700 °C annealing for sample set A while from the inset of Fig. 1(b) it is seen that the $Er^{3+}$

PL peak intensity reaches a maximum at an annealing temperature of 800 °C.

The microstructure of the samples has been investigated by TEM measurements and there are some correlations between the microstructure evolution upon annealing and the PL intensity. In the case where we observe the maximum intensity we have a large density of Ge nanoclusters for sample A and Si nanoclusters for sample B. These clusters are in an amorphous state in both cases. This essentially confirms the observations by others for Er doped Si-rich oxide [6,7] while having been reported by us before [8] for (Ge, Er) co-doped $SiO_2$. The density of clusters are reduced, are larger as the anneal temperature increases and they are transformed to nanocrystals for higher annealing temperatures. The correlation with PL is understood by regarding the absorption of excitation light by the nanoclusters provided that amorphous clusters and crystalline clusters have similar absorption properties. Both absorption and loss of Er excitation by energy back transfer could contribute to the experimental trend of PL intensity with the described microstructure related to size and distribution of nanoclusters.

We observe no luminescence in the visible range neither for sample type A nor type B. It is known that Si nanoclusters in $SiO_2$ do produce PL in the visible range with the current experimental setup [9]. Thus, the quenching of this visible luminescence indicates that the excitation energy is transferred to the Er atoms for sample set A, and the complete quenching of a strong luminescence indicates that the energy transfer process is very efficient. In the Ge case, the luminescence from Ge nanocrystals is usually weak without Er, probably due to de-excitation by defect states associated with the Ge nc-/$SiO_2$ interface or other localized defects in the close vicinity. The transfers of excitation energy to the Er atoms are thus strong; It is stronger than this parasitic energy transfer.

How strong or efficient the excitation energy transfer is from the semiconductor nanoclusters ( Si or Ge) to the Er atoms will naturally be related to the distribution of Er atoms with respect to the nanoclusters. The latter influences the physical processes for energy transfer and how strong they are. The distribution of Er could be reflected by the local environment of Er. The local environment of Er will also influence the lifetime of the excited state of Er. Thus the local environment of Er is an important factor having a large effect on the luminescence.

The shape of PL spectrum from $Er^{3+}$ will be influenced by the local environment of Er. The 4f inner-shell transition $^4I_{13/2} \rightarrow {}^4I_{15/2}$ transition of the $Er^{3+}$ which gives the PL peak around 1.5 μm is between the first exited multiplet and the ground state multiplet. The ground state and the exited state will split because of the electrical field gradient induced from neighboring atoms and thus intermixes other orbital momentum states to the f state wave function. The splitting depends upon the symmetry of the neighbors. The result is that the spectral shape of the PL peak can depend upon the site symmetry and neighbors for the $Er^{3+}$ ion. It could be commented that the PL is influenced by the environment because the atomic levels are, the 4f levels are however less influenced by the neighbor atoms, the matrix, than s- and p- valence orbitals. The latter change dramatically in comparison when their bonding environment changes. That much of the characteristics of the PL is maintained when 4f levels are involved makes $Er^{3+}$ PL a useable probe for the environment. In the case of matrixes containing oxygen it has been observed that there are two main $^4I_{13/2} \rightarrow {}^4I_{15/2}$ transitions; with peak energies around 1.550 μm and 1.538 μm [10]. These characteristics are also present in Fig.1, with a slight sample type dependant wavelength shift, which we will return to, but with the individual contributions being merged into each other and being unresolved. These main features/transitions can be resolved at lower temperatures and/or high resolution in the case of Er with Si nanocluster in $SiO_2$ [1].

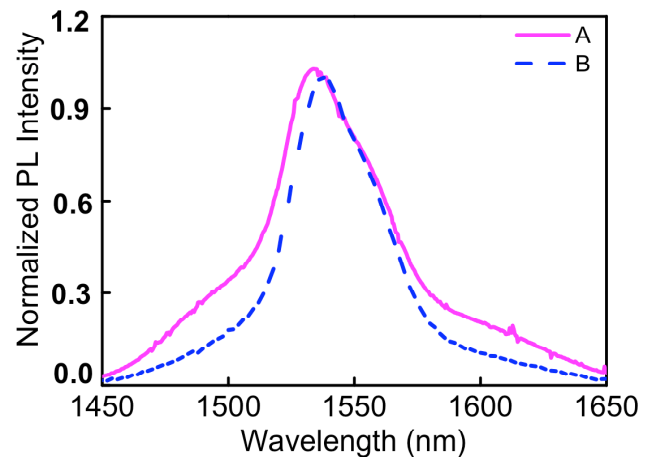

*FIG. 2 Normalized ~1.54 μm PL peak from two samples: A (Er+Ge+$SiO_2$ and B (Er+Si+$SiO_2$) for the annealing temperature which gives the maximum intensity in each case: 700 °C and 800 °C respectively.*

In Fig. 2 we compare the shape of the $Er^{3+}$ PL spectra for samples A and B for annealing temperatures when they give the maximum peak PL intensity. For this comparison the peaks have been normalized in order to notice their differences readily. In both cases the main features of the shape can be described as a peak consisting of two main unresolved contributions and a wide "socket" or "bell-rim" surrounding the main peaks. There are some statistical significant differences. These are (1) the peak positions are slightly different: ~1.534 μm for sample A (Er + Ge + $SiO_2$) and ~1.538 μm for sample B (Er + Si + $SiO_2$). (2) the width of the distribution, parameterized by the full width at half maximum (FWHM) for the two spectra is also different: 50.5 nm for sample A while for the sample it is 43 nm.(3) the broad "bell rims" at low and high wave lengths is the most striking difference in the shape of the spectra for sample A and B. These differences clearly indicates that the environment of the Er atoms are different for the different cases. From an intuitive expectation that the energy transfer would be most efficient when the Er ions are close to the nanocluster which excite them, this seems natural, and could indicate that the Er ions are surrounded with the atoms from the cluster. This is however different than the common assumption for Si, at least in the case of nanocrystals, it is assumed that the Er would be optically inactive if inside the nanocrystal [1]. The present comparison with Ge could open up for another view. In any case it indicates that the local environment for Er is different for Ge-Er co-doped $SiO_2$ than for Si-rich $SiO_2$ with Er. In both cases, the relatively wide peak is indicative of a distribution of Er atom environments; they are not all the same in either case.

We have observed that the Er and Ge are spatially correlated [8] by employing high resolution TEM with energy dispersive X-ray analysis. This is somewhat unexpected from equilibrium phase diagrams and simple chemical bond analysis, since Er-O bonds and Er-Si bonds would be stronger than Er-Ge bonds. We can speculate that Er and Ge could be correlated in the present case because the amorphous clusters containing Er may have a large configuration entropy. The differences observed in the Si and Ge cluster case is in agreement with this. On the other hand, presently it appears that the similarity in PL shapes for different annealing temperatures with the same cluster element put this explanation for the Er and Ge correlation in some doubt, if it is not kinetically controlled, which it may.

Other hyphothesises which should be addressed, or perhaps tried synthesized are amorphous clusters which may contain oxygen and Er. Thus the absorbing clusters can be optimally close to the Er ion it transfer the energy to and the environment for the Er is optimal for the first exited to ground level transition.

## 4. SUMMARY

We have compared the 1.54 µm PL from (Er, Ge) co-doped $SiO_2$ films and Er-doped Si-rich $SiO_2$ films both deposited by using magnetron sputtering a composite target in Ar plasma. The PL intensities from these two films reach their maxima after annealing the films at 700 °C and 800 °C for 30 min, respectively. The comparison of the spectral shape shows a considerable difference in the peak position and FWHM. Analysis of the experimental results confirms that amorphous Ge and Si nanoclusters play more effective role in exciting the $Er^{3+}$ luminescence than the nanocrystals.

ACKOWLEDGEMENT

Financial support from N.R.C. is appreciated.

REFERENCES

1] For examples, see review reports: A. Polman, "Erbium implanted thin film photonic materials," J. Appl. Phys. 82, 1 (1997); G. Franzò, V. Vinciguerra, and F. Priolo, "The excitation mechanism of rare-earth ions in silicon nanocrystals," Appl. Phys. A: Mater. Sci. Process. 69, 3 (1999); A. J. Kenyon, "Quantum confinement in rare-earth doped semiconductor systems," Current Opinion in Solid state & materials Science 7, 143 (2003).

[2] F. Priolo, G. Franzò, S. Coffa, A. Carnera, "Excitation and nonradiative deexcitation processes of $Er^{3+}$ in crystalline Si", Phys. Rev. B 57, 4443 (1998).

[3] J. Michel, J.L. Benton, R.F. Ferrante, D.C. Jacobsen, D.G. Eaglesham, E.A. Fitzgerald, Y.-H. Xie, J.M. Poate, L.C. Kimerling, "Impurity enhancement of the 1.54-µm $Er^{3+}$ luminescence in silicon", J. Appl. Phys. 70, 2672 (1991).

[4] A. J. Kenyon, P. F. Trwoga, M. Federighi, and C. W. Pitt, "Optical properties of PECVD erbium-doped silicon-rich silica: evidence for energy transfer between silicon microclusters and erbium ions", J. Phys.: Condens. Matter. 6, L319 (1994).

[5] C. L. Heng, T. G. Finstad, P. Storås, Y. J. Li, A. E. Gunnæs and O. Nilsen, "The 1.54-µm photoluminescence from an (Er, Ge) co-doped $SiO_2$ film deposited on Si by rf magnetron sputtering", Appl. Phys. Lett. 85, 4475 (2004).

[6] G. Franzo, S. Boninelli, D. Pacifici, F. Priolo, F. Iacona, C. Bongiorno, "Sensitizing properties of amorphous Si clusters on the 1.54-mu m luminescence of Er in Si-rich SiO2", Appl. Phys. Lett. 82, 387 (2003).

[7] C.Y. Chen, W.D. Chen, S.F. Song, Z.J. Xu, X.B. Liao, "Study on $Er^{3+}$ emission from the erbium-doped hydrogenated amorphous silicon suboxide film", J. Appl. Phys., 94, 5599 (2003).

[8] C. L. Heng, T. G. Finstad, Y. J. Li, A. E. Gunnæs, A. Olsen and P. Storås, "Ge nanoparticle formation and photoluminescence in Er doped $SiO_2$ films: influence of sputter gas and annealing", Microelectronics Journal 36, 531 (2005).

[9] J. Mayandi, T.G. Finstad, S. Foss, A. Thøgersen, U. Serincan and R. Turan, "Luminescence from silicon nanoparticles in $SiO_2$: atomic force microscopy and transmission electron microscopy studies", Phys. Scr. T126, 77 (2006).

[10] A. Janotta, M. Schmidt, R. Janssen, M. Stutzmann and Ch. Buchal, "Photoluminescence of $Er^{3+}$-implanted amorphous hydrogenated silicon", Phys. Rev. B 68, 16207 (2003).

.